\documentstyle [eqsecnum,preprint,aps,epsf]{revtex}

\makeatletter
\def\footnotesize{\@setsize\footnotesize{10.0pt}\xpt\@xpt
\abovedisplayskip 10\p@ plus2\p@ minus5\p@
\belowdisplayskip \abovedisplayskip
\abovedisplayshortskip  \z@ plus3\p@
\belowdisplayshortskip  6\p@ plus3\p@ minus3\p@
\def\@listi{\leftmargin\leftmargini
\topsep 6\p@ plus2\p@ minus2\p@\parsep 3\p@ plus2\p@ minus\p@
\itemsep \parsep}}
\long\def\@makefntext#1{\parindent 5pt\hsize\columnwidth\parskip0pt\relax
\def\strut{\vrule width0pt height0pt depth1.75pt\relax}%
$\m@th^{\@thefnmark}$#1}
\long\def\@makecaption#1#2{%
\setbox\@testboxa\hbox{\outertabfalse %
\reset@font\footnotesize\rm#1\penalty10000\hskip.5em plus.2em\ignorespaces#2}%
\setbox\@testboxb\vbox{\hsize\@capwidth
\ifdim\wd\@testboxa<\hsize %
\hbox to\hsize{\hfil\box\@testboxa\hfil}%
\else %
\footnotesize
\parindent \ifpreprintsty 1.5em \else 1em \fi
\unhbox\@testboxa\par
\fi
}%
\box\@testboxb
} %
\global\firstfigfalse
\global\firsttabfalse
\def\tabular{\let\@halignto\@empty\@tabular}
\def\endtabular{\crcr\egroup\egroup $\egroup}
\expandafter \def\csname tabular*\endcsname #1{\def\@halignto{to#1}\@tabular}
\expandafter \let \csname endtabular*\endcsname = \endtabular
\def\@tabular{\leavevmode \hbox \bgroup $\let\@acol\@tabacol
   \let\@classz\@tabclassz
   \let\@classiv\@tabclassiv \let\\\@tabularcr\@tabarray}
\def\endtable{%
\global\tableonfalse\global\outertabfalse
{\let\protect\relax\small\vskip2pt\@tablenotes\par}\xdef\@tablenotes{}%
\egroup
}%
\makeatother

\ifpreprintsty\advance\textwidth by -0.1 in\fi

\begin{document}

\def\mth	{m_{\rm th}}
\def\mphys	{m_{\rm phys}}
\def\lsim	{\,\,\vcenter{\hbox{$\buildrel{\displaystyle <}\over\sim$}}\,\,}
\def\gsim	{\,\,\vcenter{\hbox{$\buildrel{\displaystyle >}\over\sim$}}\,\,}

\preprint{UW/PT-95-09}

\title {From Quantum Field Theory to Hydrodynamics:\\
	Transport Coefficients and Effective Kinetic Theory}

\author{Sangyong Jeon\footnote
		    {%
		    Current address:
		    School of Physics and Astronomy,
		    University of Minnesota,
		    Minneapolis MN 55455
		    }
	and Laurence G.~Yaffe}

\address{Department of Physics, 
	University of Washington,
	Seattle WA 98195--1560}

\date{\today}

\maketitle

\begin{abstract}
{%
{%
\advance\leftskip -2pt
The evaluation of hydrodynamic transport coefficients
in relativistic field theory,
and the emergence of an effective kinetic theory description,
is examined.
Even in a weakly-coupled scalar field theory,
interesting subtleties arise
at high temperatures where thermal renormalization effects are important.
In this domain, a kinetic theory description in terms of the fundamental
particles ceases to be valid, but one may derive an effective kinetic theory
describing excitations with temperature dependent properties.
While the shear viscosity depends on the elastic scattering of typical
excitations whose kinetic energies are comparable to the temperature,
the bulk viscosity is sensitive to particle non-conserving
processes at small energies.
As a result, the shear and the bulk viscosities have very different
dependence on the interaction strength and temperature,
with the bulk viscosity providing an especially sensitive
test of the validity of an effective kinetic theory description.
}%

\bigskip
\bigskip

\ifpreprintsty
\thispagestyle {empty}
\newpage
\thispagestyle {empty}
\vbox to \vsize
    {%
    \vfill \baselineskip .28cm \par \font\tinyrm=cmr7 \tinyrm \noindent
    \narrower
    This report was prepared as an account of work sponsored by the
    United States Government.
    Neither the United States nor the United States Department of Energy,
    nor any of their employees, nor any of their contractors,
    subcontractors, or their employees, makes any warranty,
    express or implied, or assumes any legal liability or
    responsibility for the product or process disclosed,
    or represents that its use would not infringe privately-owned rights.
    By acceptance of this article, the publisher and/or recipient
    acknowledges the U.S.~Government's right to retain a non-exclusive,
    royalty-free license in and to any copyright covering this paper.%
    }%
\fi
}%
\end {abstract}



\section{Introduction}

 In a weakly coupled quantum field theory,
 one would expect to be able to compute
 most physical observables starting from first principles.
 However,
 at sufficiently high temperatures,
 in even the simplest scalar field theory,
 the correct evaluation of transport coefficients
 characterizing long wavelength hydrodynamic behavior is quite
 subtle.  Only recently has a thorough diagrammatic
 analysis of the bulk and shear
 viscosity appeared \cite{jeon2}, which is valid at temperatures where
 thermal renormalization effects are important.
 The purpose of this paper is to discuss the physical interpretation of
 the results of \cite{jeon2},
 and to describe the formulation of an effective
 kinetic theory which properly incorporates thermal renormalization effects
 and which generates the correct weak coupling behavior of both the bulk and
 shear viscosities.

 Existing literature in this area is somewhat sparse,
 particularly on aspects which are unique to relativistic quantum field
 theories.
 Consequently, we have tried to make the presentation reasonably self
 contained, and briefly review necessary background material.
 For simplicity, nearly all discussion will be limited to the case of a
 real scalar theory with cubic and quartic self-interactions,
 \begin{equation}
 -{\cal L} =
 {1 \over 2}
 (\partial \phi)^2
 +
 {1 \over 2} \,
 m_0^2\, \phi^2
 +
 {g\over 3!} \phi^3
 +
 {\lambda \over 4!} \phi^4
 \;,
 \label{eq:full_Lagrangian}
 \end{equation}
 with $\lambda \ll 1$, $m_0^2$ positive, and
 $g^2 = O(\lambda m_0^2)$.
 Since the theory is weakly coupled, the physical (zero temperature)
 mass of the resulting scalar particles equals $m_0$
 (after renormalization) up to radiative corrections,
 $\mphys = m_0\, (1 + O(\lambda))$.

 At non-zero temperature, the equilibrium state of this theory may be
 regarded as a fluid (or gas) of interacting spinless bosons.  For fixed
 values of the coupling constants, the pressure, energy density, and
 other thermodynamic observables depend only on the temperature.%
\footnote{%
	Since the scalar field is real, particles are their
	own antiparticles and there is no conserved number operator or
	charge to which one could couple a chemical potential.
	}
 It will be helpful to distinguish various ranges of temperature:
 \begin{itemize}
 \item[{\it i})]
 	$0 < T \ll \mphys$.  The system is non-relativistic and dilute.
 	The equilibrium particle density is exponentially small,
	$n \sim (\mphys T)^{3/2} e^{-\mphys/T}$.

 \item[{\it ii})]
	$\mphys \lsim T \ll \mphys/\sqrt{\lambda}$.  The system is
	relativistic, but thermal corrections to the effective
	particle mass (or scattering amplitudes) are negligible.

 \item[{\it iii})]
	$T \approx \mphys/\sqrt{\lambda}$.  The thermal correction to
	the particle mass, of order $\sqrt{\lambda}\,T$, is comparable to
	the zero temperature mass.  The system may no longer be regarded
	as a weakly interacting collection of the underlying fundamental
	particles.

 \item[{\it iv})]
 	$\mphys/\sqrt{\lambda} \ll T \ll \mphys/\lambda$.
	The zero temperature mass is
 	negligible compared to the thermal mass shift, but the zero
	temperature mass still dominates the trace anomaly
	of $\beta(\lambda)\phi^4 /4!$
	in $\langle {T_\mu}^\mu \rangle$.
 \item[{\it v})]
	$T \gg \mphys/\lambda$.  The zero temperature mass is negligible
	even in $\langle {T_\mu}^\mu \rangle$.
 \end{itemize}

    The most interesting domains, from a theoretical perspective,
are the high temperature ranges ({\it iii--v}) where thermal
renormalization effects are important.
Table \ref {tablea} summarizes the qualitative behavior of various
quantities at these temperatures.

 \begin{table}
    \begin {center}
    \tabcolsep=8pt
    \renewcommand{\arraystretch}{0.75}
    \begin {tabular}{|l@{\kern 30pt}l|}             \hline
    particle density
	& $\llap {$n$} = O(T^3)$
	\\
    energy density
	& $\llap {$\varepsilon$} = O(T^4)$
	\\
    effective particle mass
	& $\llap {$\mth$} = O(\lambda^{1/2} \, T)$
	\\
    on-shell self energy
	& $\llap {$\Sigma(p)$} = O(\lambda \, T^2)
			    + i \, O(\lambda^2 \, T^2)$
	\\
    thermal width \ \ ($p = O(\mth)$)
	& $\llap {$\Gamma_p$} = O(\lambda^{3/2} \, T)$
	\\
    thermal width \ \ ($p = O(T)$)
	& $\llap {$\Gamma_p$} = O(\lambda^2 \, T)$
	\\
    mean free time \ ($p = O(T)$)
	& $\llap {$\tau_{\rm f}$} = O(\lambda^{-2} \, T^{-1})$
	\\
    elastic cross section ($p = O(T)$)
	& $\llap {$\sigma$} = O(\lambda^2 \, T^{-2})$
	\\
    speed of sound
	& $\llap {$v_{\rm s}$} = 1/\sqrt 3 + O(\lambda)$
	\\
    shear viscosity
	& $\llap {$\eta$} = O(\lambda^{-2} \, T^3)$
	\\
    bulk viscosity ($T \gsim O(\mphys / \lambda)$)
	& $\llap {$\zeta$} = O(\lambda \, T^3 \ln^2\lambda)$
	\\
    bulk viscosity ($T = O(\mphys / \sqrt{\lambda})$)
	& $\llap {$\zeta$} = O(\mphys^4\, T^{-1} \,
			    \lambda^{-5/2}\ln^2\lambda)$
	\\
    \hline
    \end {tabular}
    \end {center}
    \caption
    {%
    \label {tablea}
    Scaling behavior of various quantities in
    high temperature
    scalar field theory.
    The estimates hold in the domain
    $T \gsim \mphys / \protect\sqrt {\lambda}$
    where the one-loop thermal contribution dominates the (real part of the)
    single particle self energy $\Sigma(p)$.
    If the scalar field has only quartic interactions, then
    the last result for the bulk viscosity acquires an additional
    factor of $\lambda^{-1/2}$.
    See section \protect\ref {results} for more detailed expressions.
    }%
 \end{table}

\noindent
The scaling behavior of the effective particle mass and
thermal width will be essential ingredients in the following
discussion.
The thermal width is the inverse of the mean free time between scattering
(up to statistical factors) and equals the displacement of the single particle
pole away from the real frequency axis.
The size of the thermal width follows directly from
the imaginary part of the on-shell self energy
divided by the particle energy.
For weak coupling, the thermal width is small compared to the
effective mass because the imaginary part of the on-shell self energy
first arises from two-loop graphs, whereas the real part has one-loop
contributions.
The results displayed for the shear and bulk viscosities
will be discussed in detail in section \ref {results}.

\section {Transport Coefficients and Basic Kinetic Theory}

 In a fluid with no conserved particle number,
 the stress-energy tensor $T_{\mu\nu}$ is
 the only locally conserved current,
 and fluctuations in the energy and momentum densities
 are the only hydrodynamic modes.\footnote{%
	These are fluctuations whose relaxation time diverges as the
	wavelength of the fluctuation increases.  Such
	fluctuations determine the behavior of the system at
	arbitrarily long times and large distances.
	}
 Two transport coefficients, the shear and bulk viscosities,
 (denoted $\eta$ and $\zeta$, respectively)
 characterize the resulting hydrodynamic response.%
\footnote{%
	Because there is no conserved particle number, thermal
	conductivity is not an independent transport coefficient.
	}
 If the system is slightly perturbed from equilibrium, then the
 non-equilibrium expectation of $T_{\mu\nu}$ will satisfy the
 constitutive relation (in a local fluid rest frame),
 \begin{equation}
 \langle T_{ij}\rangle
 =
 \delta_{ij} \langle {\cal P} \rangle
 -{\eta \over \langle \varepsilon {+} {\cal P} \rangle}
     \left(
 	 \nabla_i\langle {T^0}_j \rangle
       + \nabla_j \langle {T^0}_i \rangle
       - {\textstyle {2 \over 3}} \delta_{ij}
 	\nabla^l \langle {T^0}_l \rangle
 	\right)
       -{\zeta \over \langle \varepsilon {+} {\cal P} \rangle}
        \delta_{ij}  \nabla^l \langle {T^0}_l \rangle
 \;,
 \label{eq:constitutive_rel}
 \end{equation}
 together with the exact conservation law,
 $\partial_{\mu}\langle T^{\mu\nu} \rangle = 0$.
 Here $T_{ij}$ is the spatial part of the stress-energy tensor,
 $\varepsilon \equiv T_{00}$ is the energy density,
 and $\langle {\cal P} \rangle$ is the local equilibrium pressure.
 The constitutive relation (\ref{eq:constitutive_rel})
 is valid for small fluctuations
 in the limit in which the scale of the variation in
 $\langle T_{\mu\nu} \rangle$ is arbitrarily large compared to
 microscopic length scales (such as the mean free path of excitations).

 The shear viscosity $\eta$ characterizes the diffusive relaxation of
 transverse momentum density fluctuations;
 $\eta/\langle \varepsilon{+}{\cal P} \rangle$ is the diffusion constant
 for such shear fluctuations.
 The bulk viscosity $\zeta$ characterizes the departure from
 equilibrium during a uniform expansion.
 If the divergence of the fluid flow is constant,
 then the pressure differs from the local equilibrium value by the
 bulk viscosity times the expansion rate, or
 $\zeta \, \nabla^i \langle T^0_i\rangle
 / \langle \varepsilon {+} {\cal P} \rangle $.
 Both shear and bulk viscosity contribute to the attenuation of sound waves;
 the decay rate of sound waves with wavenumber $\bf k$ is
 ${\bf k}^2 (\textstyle {4 \over 3} \, \eta {+} \zeta) \,
  /\langle \varepsilon{+}{\cal P} \rangle$ \cite{Groot}.
 The bulk viscosity vanishes identically in a scale invariant theory
 \cite {Weinberg}.
 This follows from the vanishing trace of the stress-energy tensor,
 ${T_\mu}^\mu = 0$, in any scale invariant theory,
 and reflects the fact that a uniform dilation of an equilibrium
 distribution function remains in equilibrium (at a modified temperature)
 if the dispersion relation is scale invariant.

     Transport coefficients are proportional to the mean free path
 of the scattering processes responsible for relaxation of the
 associated hydrodynamic modes.
 The shear viscosity is proportional to the two body elastic
 scattering mean free path.
 In (scale non-invariant) relativistic theories,
 the bulk viscosity is proportional to the mean free path
 for particle number changing processes.
 This may be understood by noting that after a uniform expansion,
 a change in the total number of particles is required
 in order to re-equilibrate at a different temperature.%
 \footnote
    {%
    Under uniform expansion, a non-relativistic gas of molecules
    relaxes by converting internal energy (vibrational or rotational)
    into kinetic energy.
    In contrast, a relativistic gas of structureless particles
    relaxes by converting rest-mass energy into kinetic energy.
    }
 Note that decreasing the interaction strength will increase
 the mean free paths and thus normally increase transport coefficients.
 Consequently, the weak coupling expansion of viscosities
 typically begins with negative powers of the coupling.

 Most textbook discussions of the evaluation of transport coefficients
 (such as \cite{Groot}) begin by assuming the validity of a kinetic theory
 description of the interacting fluid.
 One argues that the system may be characterized by a distribution
 function $f(x, p)$ giving the phase space probability density of the
 fundamental particles comprising the fluid.
 Although written as if it depends on an arbitrary four momentum, the
 distribution function is only defined for on-shell particles, for which
 $p^0 = E_p \equiv \sqrt{{\bf p}^2{+}\mphys^2}$.  The time dependence
 of the distribution function is governed by a Boltzmann equation,
 \begin{equation}
 {p^\mu \over E_p}{\partial \over \partial x^\mu}\, f(x,p)
 =
 {\textstyle{1\over 2}}
 \int_{123} d\Gamma_{12\leftrightarrow 3p}
 \bigg(
 f_1\,f_2\,(1{+}f_3)\,(1{+}f_p)
 -
 (1{+}f_1)\,(1{+}f_2)\,f_3\,f_p
 \bigg)
 \label{eq:Boltzmann}
 \;,
 \end{equation}
 where $d\Gamma_{12\leftrightarrow 3p}$ is the differential transition rate
 for particles of momenta $k_1$ and $k_2$ to scatter into momenta $k_3$ and
 $p$,
 \begin{equation}
     d\Gamma_{12\leftrightarrow 3p}
     \equiv
     {1\over 2 E_p}\,
     \Big| {\cal T}(p,k_3;k_2,k_1) \Big|^2 \,
     \prod_{i=1}^3 {d^3 {\bf k}_i \over (2\pi)^3 (2E_{k_i})}\,
     (2\pi)^4 \delta (k_1{+}k_2{-}k_3{-}p) \,
     \;,
 \label {diff-trans-rate-1}
 \end{equation}
 and $f_i \equiv f(x, k_i)$, $f_p \equiv f(x, p)$.
 The collision term (or the right hand side of (\ref{eq:Boltzmann}))
 vanishes when the distribution function is an equilibrium Bose
 distribution with an (inverse) temperature $\beta$ and flow velocity
 $u^\mu$, or
 \begin{equation}
 f^{\rm eq}_\beta (x,p)
 = n(\left| u^\mu p_\mu\right|) \,,
 \end{equation}
 with $n(E)$ the usual Bose distribution function at inverse
 temperature $\beta$,
 \begin {equation}
    n(E) \equiv \left( e^{\beta E} - 1 \right)^{-1} \,.
 \end {equation}

 To extract transport coefficients, it is sufficient to consider
 perturbations away from equilibrium which are arbitrarily small and
 slowly varying.  Writing the distribution function as a local
 equilibrium piece plus a non-equilibrium correction,
 \begin{equation}
 f(x,p)
 = f^{\rm eq}_{\beta(x)}(x,p)\,
 \left\{
 1 - \chi(x,p) \, [1+f^{\rm eq}_{\beta(x)}(x,p)]
 \right\}
 \;,
 \label{eq:f_plus_df}
 \end{equation}
 one may linearize the Boltzmann equation and expand in powers of
 $\nabla u$ or $\nabla \beta$.  After using the conservation relation,
 $\partial_\mu T^{\mu\nu}=0$, to express time derivatives in terms of
 spatial gradients,\footnote{%
	And imposing the Landau-Lifshitz condition
	$T^{\mu\nu} u_\nu = T^{\mu\nu}_{\rm eq} u_\nu$ to make the
	decomposition (\ref{eq:f_plus_df}) unique.
	}
 one finds that (in the fluid rest frame at a particular point $x$)
 \cite{Groot},
 \begin{equation}
 \chi(x,p)
 =
 \beta(x)\, A(x,p)\, \nabla {\cdot} {\bf u}(x)
 +
 \beta(x)\, B(x,p)
 \left[\,
 \hat{\bf p} {\cdot} \nabla ({\bf u}(x) {\cdot} \hat{\bf p})
 -
 {\textstyle {1\over 3}} \nabla {\cdot} {\bf u}(x)
 \right] ,
 \label {phi-from-A&B}
 \end{equation}
 where the coefficient $B$ multiplying the shear in the flow
 satisfies the linear inhomogeneous integral equation
 \begin{eqnarray}
 p_i p_j - {\textstyle{1\over 3}} {\bf p}^2 \delta_{ij}
 &=& \displaystyle
 {E_p \over 2}
 \int_{123} d\Gamma_{12\leftrightarrow 3p}\,
 (1{+}n_1)\,(1{+}n_2)\,n_3 (1{+}n_p)^{-1}\,
 \nonumber
 \\
 & & \displaystyle \qquad {}\times
 \bigg[
 B_{ij}(p) + B_{ij}(k_3) - B_{ij}(k_2) - B_{ij}(k_1)
 \bigg] ,
 \label{eq:kinetic_shear_eq}
 \end{eqnarray}
 with $B_{ij}(p) \equiv
 B(p)(\hat{p}_i \hat{p}_j - {\textstyle{1\over 3}} \delta_{ij})$,
 and all quantities evaluated at the point $x$.
 The coefficient $A$ multiplying the divergence of the flow satisfies an
 analogous integral equation,
 \begin{eqnarray}
 {\textstyle{1\over 3}} {\bf p}^2
 -
 v_{\rm s}^2 \, ({\bf p}^2{+}\mphys^2)
 &=& \displaystyle
 {E_p \over 2}
 \int_{123} d\Gamma_{12\leftrightarrow 3p}\,
 (1{+}n_1)\,(1{+}n_2)\,n_3 (1{+}n_p)^{-1}\,
 \nonumber
 \\
 & & \displaystyle \qquad {}\times
 \bigg[
 A(p) + A(k_3) - A(k_2) - A(k_1)
 \bigg] ,
 \label{eq:kinetic_bulk_eq}
 \end{eqnarray}
 with $v_{\rm s} \equiv (\partial {\cal P}/\partial \varepsilon)^{1/2}$
 the (local equilibrium) speed of sound,
 together with the constraint
 \begin {equation}
    0 = \int {d^3p \over (2\pi)^3} \> E_p \, n(E_p) \, (1 + n(E_p)) \, A(p) \,.
 \end {equation}
 Finally, inserting the distribution function into the kinetic theory
 stress-energy tensor,
 \begin{equation}
 T^{\mu\nu}(x)
 =
 \int {d^3 {\bf p}\over (2\pi)^3 E_p}\,\,
 p^\mu p^\nu f(x,p)
 \;,
 \end{equation}
 and comparing with the constitutive relation
 (\ref{eq:constitutive_rel}) yields
 \begin{mathletters}
 \begin{eqnarray}
 & \displaystyle \eta =
 {\beta \over 15}
 \int {d^3 {\bf p}\over (2\pi)^3 E_p}\,
 {\bf p}^2\, n(E_p)\, (1+n(E_p))\, B({\bf p})
 \;,
 & \displaystyle
 \label{eq:kinetic_shear}
 \\
 \noalign{\hbox{and}}
 & \displaystyle
 \zeta =
 \beta
 \int {d^3 {\bf p}\over (2\pi)^3 E_p}\,
 \left(
 {\textstyle{1\over 3}}{\bf p}^2
 {-}
 v_{\rm s}^2 ({\bf p}^2 {+} \mphys^2)
 \right)\,
 n(E_p)\, (1+n(E_p))\, A({\bf p})
 \;.
 & \displaystyle
 \label{eq:kinetic_bulk}
 \end{eqnarray}
 \label{eq:kinetic_viscosities}%
 \end{mathletters}%
 \noindent
 Hence,
 in this kinetic theory treatment,
 a quantitative evaluation of viscosities
 requires solving the linear integral equations
 (\ref{eq:kinetic_shear_eq}) and
 (\ref{eq:kinetic_bulk_eq}) and then computing the final momentum
 integrals in (\ref{eq:kinetic_viscosities}).
 These results
 (\ref{eq:kinetic_bulk_eq}--\ref{eq:kinetic_viscosities})
 for the viscosities can only trusted within the domain of validity of
 the underlying Boltzmann equation (\ref{eq:Boltzmann}).  Basic
 assumptions underlying kinetic theory which must hold include the
 following.
 \begin{itemize}
 \item[{\it a})] The collision time is negligible compared to the the mean
 free time between collisions of the fundamental particles.

 \item[{\it b})] Between collisions, particles may be regarded as propagating
 classically with definite momentum and energy.

 \item[{\it c})] The on-shell energy and momentum of particles in between
 collisions satisfy the zero temperature free particle dispersion
 relation, $E_p = \sqrt{{\bf p}^2{+}\mphys^2}$.
 \end{itemize}

 When the system is non-relativistic, $T \ll  \mphys$, the mean free
 time is exponentially large (compared to the Compton time
 $\hbar/\mphys c^2$) and the above assumptions are well satisfied.  In
 the relativistic domain, $T \gsim \mphys$, the situation is more
 involved.  In this regime, the density of particles scales as $T^3$, a
 typical two-body elastic scattering cross section is
 $\sigma \sim \lambda^2/T^2$, and so the mean free time is
 $O(1/\lambda^2 T)$
 (or $O (1/\lambda^{3/2} T)$ for soft particles due to
 Bose-enhanced stimulated emission).
 In contrast, the typical collision time
 (determined by the variation of the phase shift with energy) is
 $O(\lambda^2/T)$; hence condition ($a$) is satisfied as long as the
 theory is weakly coupled.  Quantum uncertainties in the energy or
 momentum of a particle propagating between collisions are negligible
 provided the kinetic energy times the mean free time is large compared
 to $\hbar$.  For particles with typical $O(T)$ energies, this
 condition again merely requires $\lambda \ll 1$.  But since the mean
 free time becomes arbitrarily small as the temperature increases,
 ``soft'' particles with momentum of order of their rest mass cannot
 be viewed as propagating classically when $T \gsim \mphys/\lambda^2$.
 Moreover, standard kinetic theory fails long before this temperature
 is reached due to condition ($c$).  The collision term in the Boltzmann
 equation summarizes the effects of scatterings in which particles change
 their momenta in a near-random manner which may be regarded as
 destroying phase coherence.  It does {\em not\/} describe the coherent
 change in phase caused by exactly forward scattering.  The amplitude
 for a soft particle to propagate through the surrounding medium will be
 modified due to phase shifts arising from forward scattering
 interactions, and this will change the dispersion relation from the
 zero temperature form.\footnote{%
	This, of course, is exactly how the index of refraction for
	light is generated.
	}
 For a hot scalar theory, this is precisely the origin of the well-known
 thermal correction to the effective particle mass,
 \begin{equation}
 \mth(T)^2
 =
 \mphys^2
 +
 {\lambda \, T^2 \over 24}
 \times
 \left(1 + O \! \left({\mphys \over T}\right) \right)
 .
 \label{eq:thermal_mass}
 \end{equation}
 For simplicity, contributions arising from the cubic coupling have
 not been displayed.
 Here (and henceforth) $\lambda$ and $g^2$ are renormalized couplings
 evaluated at the scale $T$.%
 \footnote{%
	A renormalization point of order $T$ is needed to
	avoid large logarithms in higher order corrections.
	}
 The thermal mass correction is
 negligible when $T\lsim O(\mphys)$, but when
 $T \gsim O(\mphys/\sqrt{\lambda})$ the mass correction is
 significant and the standard Boltzmann equation fails to
 describe correctly the propagation of particles with soft $O(\mphys)$
 momenta.  It is important to note that forward scattering effects will
 also change the effective cross sections of soft particles propagating
 through the medium.

 When $T \gg \mphys$, one might expect the inapplicability of the Boltzmann
 equation for soft particles to be irrelevant, since
 particles with $O(\mphys)$ momenta comprise a small
 $O((\mphys/T)^2)$ fraction of the total.
 This is true for some physical observables (including
 thermodynamic quantities such as the pressure or energy density, and
 also the shear viscosity).  However, as will be seen explicitly below,
 the bulk viscosity is predominately sensitive to soft $O(\mphys)$
 momenta.\footnote{%
	This difference between the shear and bulk viscosity is easy to
	see from equations
 	(\ref{eq:kinetic_shear_eq}--\ref{eq:kinetic_viscosities}).
	The factors of ${\bf p}^2$ in the shear viscosity integrand
	(\ref{eq:kinetic_shear}) and the inhomogeneous term in
	(\ref{eq:kinetic_shear_eq}) combine to suppress the contribution
	of soft
	momenta by four powers of $|{\bf p}|$ relative to the case of
	the bulk viscosity.
	}

 There is an additional problem with the kinetic theory treatment for
 the bulk viscosity.  The integral equation
 (\ref{eq:kinetic_bulk_eq}) has no solution!
 The kernel of the equation has a zero mode (which is not
 orthogonal to the source term).  The zero mode is a consequence of the
 conservation of particle number.  The original
 $g\phi^3{+}\lambda\phi^4$ theory does not have a conserved particle
 number, but the Boltzmann equation (\ref{eq:Boltzmann}) with only
 two-particle scattering terms included obviously does conserve the
 number of particles.
 As stated earlier, the bulk viscosity, which characterizes the
 relaxation of the system after a uniform expansion, is directly
 sensitive to particle number changing processes since a (scale
 non-invariant) system undergoing uniform expansion cannot re-equilibrate
 without changing the number of particles.\footnote{%
	If the number of particles is conserved (as in a
	non-relativistic field theory), equilibrium states will
	depend on a chemical potential as well as the temperature, and
	the zero mode in (\ref{eq:kinetic_bulk_eq}) will be removed by the
	additional subsidiary condition on $\chi$ needed to make the
	local chemical potential uniquely defined.  In such a theory, a
	uniform expansion will, of course, produce a change in the
	chemical potential.
	}
 Consequently, higher order particle number changing terms must be
 included in the Boltzmann equation (\ref{eq:Boltzmann}) even though
 they are suppressed by additional powers of $\lambda$.  This will be
 described more explicitly below.

 In summary, standard kinetic theory (with number changing processes
 included) is adequate for calculating transport coefficients in a
 weakly coupled theory in the temperature regimes where
 $T\ll \mphys/\sqrt{\lambda}$, but not in the high temperature regimes
 with $T\gsim \mphys/\sqrt{\lambda}$.
 In order to derive transport coefficients in this domain,
 one should start directly from the underlying field theory.

\section {Diagrammatic Evaluation of Transport Coefficients}

 The shear and bulk viscosities may be extracted from the zero momentum,
 small frequency limit of the spectral density of the equilibrium
 stress tensor--stress tensor correlation function.  One finds that
 \cite{jeon2}
 \begin{mathletters}
 \begin{eqnarray}
 \eta
 &=& \displaystyle
 {1\over 20}\,
 \lim_{\omega \to 0}\, {1\over \omega}\,
 \int d^4 x\, e^{i\omega t}\,
 \langle [{\pi}_{lm}(t,{\bf x}), {\pi}_{lm}(0)] \rangle_{\rm eq}
 \label{eq:kubo_shear}
 \;,
 \\
 \noalign{\hbox{and}}
 \zeta
 &=& \displaystyle
 {1\over 2}\,
 \lim_{\omega \to 0}\, {1\over \omega}\,
 \int d^4 x\, e^{i\omega t}\,
 \langle [\bar{\cal P}(t,{\bf x}), \bar{\cal P}(0)] \rangle_{\rm eq}
 \;.
 \label{eq:kubo_bulk}
 \end{eqnarray}
 \label{eq:kubo_relations}%
 \end{mathletters}%
 \noindent
 Here,
 $
 {\pi}_{lm} \equiv
 \nabla_l {\phi} \, \nabla_m {\phi}
 -{\textstyle {1\over 3}}
 \delta_{lm} (\nabla {\phi})^2 
 $
 is the traceless part of the stress tensor and
 $\bar{\cal P} \equiv {\cal P}{-}v_{\rm s}^2\varepsilon$
 is the pressure minus the energy density
 times the square of the speed of sound.\footnote{%
	The Kubo relation (\protect{\ref{eq:kubo_relations}}b)
	is equally correct if the pressure ${\cal P}$ is used in place of
	$\bar{\cal P}$, since commutators of the energy density vanish
	at zero momentum (due to energy conservation).  However, as
	shown in \cite{jeon2} and explained below, the particular choice of
	$\bar{\cal P}$ given is appropriate for deriving an ``effective''
	kinetic theory description.
	}
 These Kubo relations
 provide the natural starting point for a field theory evaluation.
 The spectral density equals the discontinuity in the (Fourier
 transformed) stress-stress correlation function and has a perturbative
 expansion generated by the sum of cut diagrams with two insertions of
 $T_{\mu\nu}$ \cite{jeon1}.
 Naively, one would expect the leading weak coupling contribution to
 arise solely from the single one-loop diagram shown in Fig.~1.

\begin{figure}
 \setlength {\unitlength}{1cm}
\vbox
    {%
    \begin {center}
 \begin{picture}(0,0)
 \end{picture}
	\leavevmode
	\def\epsfsize	#1#2{0.3#1}
	\epsfbox {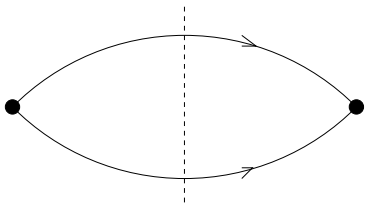}
    \end {center}
    \caption{%
 The cut one loop diagram contribution to the viscosity.
 }
 \label{fig:one_loop}
    }%
\end {figure}
 \noindent
 This is correct for generic values of the external 4-momentum, but is
 completely incorrect in the limit of vanishing external momentum
 and frequency.  Finite temperature propagators have poles (in
 frequency) with both $+i\epsilon$ and $-i\epsilon$
 prescriptions.
 When the external 4-momentum vanishes, the product of propagators
 corresponding to the graph in Fig.~1 contains terms in which the contour of
 the frequency integration is pinched between coalescing poles, thereby
 producing an on-shell divergence.
 As always, such divergences have a simple physical origin
 \cite{Coleman}.  When a small momentum is introduced by an insertion of
 $T_{\mu\nu}$, an on-shell (bare) particle in the thermal medium
 can absorb the external momentum and become slightly off-shell.
 The amplitude is proportional to the length of time the particle can
 remain off-shell.  As the external momentum vanishes,
 the virtual particle moves on-shell and the integral over the
 propagation time diverges.

 However, at non-zero temperature, no excitation can actually propagate
 indefinitely through the thermal medium without suffering collisions off
 other excitations.  In a scalar field theory, a single particle
 excitation of momentum ${\bf k}$ acquires a finite lifetime $\tau_k$,
 or non-zero thermal width $\Gamma_k \equiv 1/\tau_k$, due to
 the $O(\lambda^2)$ imaginary part of the on-shell two-loop
 self-energy.  To examine the limit of vanishing external momentum, one
 must resum the single particle self-energy insertions which will shift
 the poles in the single particle propagator from
 $\pm E_k^0\,{\pm}\,i\epsilon$ to $\pm E_k^{\rm th}\,{\pm}\,i\Gamma_k$
 (where $E_k^{\rm th} \equiv \sqrt{{\bf k}^2{+}m(T)^2}$).
 This serves to regulate the apparent on-shell singularity, and makes
 the one-loop diagram in Fig.~1 yield a finite result proportional to the
 single particle lifetime.  However, since the lifetime is
 $O(1/\lambda^2)$ (for particles with $O(T)$ momenta)
 this means that higher loop diagrams can be just
 as important as the one-loop contribution if they are sufficiently
 infrared sensitive.

\begin{figure}
 \setlength {\unitlength}{1cm}
\vbox
    {%
    \begin {center}
 \begin{picture}(0,0)
 \end{picture}
	\leavevmode
	\def\epsfsize	#1#2{0.4#1}
	\epsfbox {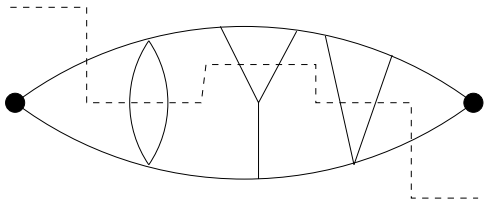}
    \end {center}
    \caption{%
 A typical cut ladder diagram for the shear viscosity
 in $g\phi^3{+}\lambda\phi^4$ theory
 containing $O(\lambda^2)$, $O(g^2\lambda)$, and $O(g^4)$
 ``rungs''.
 }
 \label{fig:typical}
    }%
\end {figure}

 For the shear viscosity, a careful analysis shows that one must sum
 all cut ``ladder-like'' diagrams of the type illustrated in Fig.~2.  See
 \cite{jeon2} for details.  This is similar to the situation in
 non-relativistic systems \cite{Mahan},
 except that instead of having ladders built
 from an instantaneous two body interaction, one must deal with
 ladder graphs containing far more complicated ``rungs''.  Nevertheless,
 one may formally sum all cut ladder-like graphs by introducing
 an effective vertex ${\cal D}_{\pi}(k,q{-}k)$ satisfying a linear
 equation of the form
 \begin{equation}
 {\cal D}_{\pi}(k,q{-}k)
 =
 {\cal I}_{\pi}(k,q{-}k)
 +
 \int {d^4 p \over (2\pi)^4} \,
 {\cal M}(k{-}p) \,
 {\cal F}(p,q{-}p)
 {\cal D}_{\pi}(p,q{-}p)
 \;.
 \label{eq:vertex_eq4}
 \end{equation}
 The effective vertex actually has a four components (in order to
 represent the four different choices for which legs are above and below
 the cut), while
 ${\cal M}(k{-}p)$ and ${\cal F}(p,q{-}p)$ are $4\times 4$ matrices
 representing the rungs and side-rails of the ladder, respectively.
 These matrices
 have entries consisting of various products of cut and uncut
 propagators.
 The inhomogeneous term ${\cal I}_\pi(k,q{-}k)$ represents the vertex
 factors corresponding to an insertion of the traceless stress tensor.
 The explicit form of each of these quantities may be found in \cite{jeon2}.
 Closing the two legs of the effective vertex with a second insertion of
 the traceless stress tensor produces the sum of all ladder-like graphs
 contributing to the shear viscosity, so that
 \begin{equation}
 \eta = {\beta \over 10}\,
 \lim_{q^0 \to 0}\lim_{{\bf q}\to 0}\,
 \int{d^4 k\over (2\pi)^4}\,\,
 {\cal I}_{\pi}(k,q{-}k){\cdot}{\cal F}(k,q{-}k)\, {\cal D}_{\pi}(k,q{-}k)
 \times (1 + O(\sqrt{\lambda}))
 \;.
 \label{eq:shear_ladder_sum}
 \end{equation}

 In the limit of vanishing external momentum $q$,
 one may perform the frequency integration and extract the leading order
 behavior from the nearly pinching-pole contributions \cite{jeon2}.
 Moreover, by using the finite temperature optical theorem
 \cite{jeon2,Kobes} the $4\times 4$ kernel ${\cal MF}$ may be shown
 to equal
 a rank one matrix (up to corrections subleading in $\lambda$), thereby
 allowing one to reduce the equation to a single component, three
 dimensional integral equation.  The result is identical to equation
 (\ref{eq:kinetic_shear_eq}) for the spin-two part of the
 Boltzmann distribution function, and (\ref{eq:kinetic_shear})
 for the kinetic theory shear
 viscosity provided one:
 \begin{itemize}
 \item[{\it a})] identifies the shear response
 $B(k)$ with the effective vertex divided by the
 imaginary part of the single particle self-energy,
 \item[{\it b})] uses the thermal mass $m_{\rm th}(T)$ instead of the
 zero-temperature mass in the dispersion relation defining on-shell
 momenta, and
 \item[{\it c})] uses an effective temperature-dependent
 ``scattering amplitude'' equal to the usual tree-level
 amplitude but evaluated with finite temperature retarded propagators,%
 \footnote{%
	Since the intermediate propagator in
	(\protect{\ref{eq:almost_scattering_amp}}) cannot go on-shell,
	the $i\epsilon$ prescription in the retarded propagator is
	actually irrelevant.  Note however, that using the real time
	Feynman propagator in the scattering amplitude is incorrect as
	this differs (off-shell) by a $(1{+}n(E_p))$ Bose distribution
	factor.
	}%
 \begin{equation}
     {\cal T}(p_1,p_2;p_3,p_4)
     \equiv
     \lambda
     - \bar g^2
     \left( G_R(p_1{+}p_2) +G_R(p_1{-}p_4) +G_R(p_1{-}p_3) \right)
     \;,
 \label{eq:almost_scattering_amp}
 \end{equation}
 where
 $G_R(p_1{+}p_2) = [(p_1{+}p_2)^2 + \mth^2]^{-1}$
 and $\bar g = g + \lambda \langle \phi \rangle$
 is the ``shifted'' cubic coupling constant that
 results when one shifts the field by its thermal
 expectation value $\langle \phi \rangle$
 in order to remove tadpole diagrams.%
 \footnote
    {%
    Scattering amplitudes, strictly speaking, do not exist at non-zero
    temperature, since all excitations have finite lifetimes.
    However, in this weakly coupled theory,
    the effective scattering amplitude
    (\ref{eq:almost_scattering_amp})
    provides a meaningful characterization of scattering processes which
    occur on time scales short compared to the single particle lifetime.
    }
 \end{itemize}

 The calculation of bulk viscosity requires considerably
 more care than the shear viscosity.
 In addition to ladder diagrams of the type shown in Fig.~2,
 one must also sum diagrams containing iterations of higher order
 number-changing scattering processes, and include thermal ``vertex
 renormalization'' subgraphs \cite{jeon2}.  Examples are shown in Fig.~3.
\begin{figure}
 \setlength {\unitlength}{1cm}
\vbox
    {%
    \begin {center}
 \begin{picture}(0,0)
 \end{picture}
	\leavevmode
	\def\epsfsize	#1#2{0.4#1}
	\epsfbox {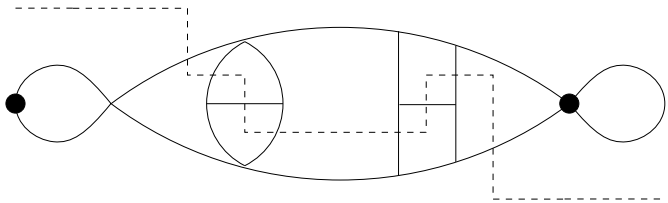}
    \end {center}
    \caption{%
 A typical graph containing $O(g^2\lambda^2)$ and
 $O(g^6)$ two-to-three particle ``rungs'', plus ``thermal
 renormalization'' of the stress tensor vertices.  Graphs such as this
 contribute to the leading order weak coupling behavior of the bulk
 viscosity.
 }
 \label{fig:typical_2}
    }%
\end {figure}
\noindent
 Nevertheless, one may again sum all the relevant diagrams by
 introducing an effective vertex ${\cal D}_{\bar{\cal P}}(k,q{-}k)$
 satisfying a linear equation of the same form as in
 (\ref{eq:vertex_eq4}).  The appropriate kernel now contains the
 previous $O(\lambda^2)$ subdiagrams plus $O(g^2\lambda^2)$
 number changing subdiagrams.\footnote{%
	Or, for a pure $\lambda\phi^4$ theory, $O(\lambda^4)$
	two-to-four particle subdiagrams.
	}
 The inhomogeneous term receives $O(\lambda)$ corrections
 involving the one-loop contributions to the thermal mass and the speed
 of sound.
 These vertex corrections cannot be neglected at high temperatures
 because the speed of sound (squared) approaches 1/3, producing a cancellation
 in the leading $O({\bf p}^2)$ part of the inhomogeneous term
 (\ref {eq:kinetic_bulk_eq}).
 Consequently,
 an insertion of $\bar {\cal P}$
 (or pressure minus $v_{\rm s}^2$ times the energy density)
 is $O(\mphys^2)$,
 even when the loop momentum is of $O(T)$.
 Hence,
 vertex corrections which are $O (\lambda T^2)$
 can be comparable to
 the zeroth order term.

 Once again, one may perform the frequency integrations and extract the
 leading behavior from the nearly pinching-pole
 contributions,\footnote{%
 	Subleading non-pinching pole terms in the kernel can be
	neglecting only if the inhomogeneous term is orthogonal to the
	zero modes of the reduced pinching-pole kernel (as well as
	orthogonal to the zero modes of the full kernel).  Imposing this
	condition forces
	the energy density coefficient in the source
	$\bar{\cal P} = {\cal P}{-}v_{\rm s}^2\varepsilon$ to equal the
	speed of sound (including one-loop corrections) \cite{jeon2}.
	}
 show that the resulting kernel is dominated by a rank one matrix, and
 reduce the equation to a single component, three dimensional
 integral equation.  This has the same form as
 Eq.~(\ref{eq:kinetic_bulk_eq})
 (with $A(p)$ identified with
 the effective vertex divided by the imaginary part of the self-energy)
 except that:
 \begin{itemize}
 \item[{\it a})] In addition to the two particle elastic scattering
 term, the right hand side now contains a particle number
 changing term proportional to the square of the tree level
 two-to-three particle ``scattering amplitude'',\footnote{%
	Or, for a pure $\lambda\phi^4$ theory, the two-to-four
	particle amplitude
 	$
	{\cal T}_{\Delta N}
	=
	-i \lambda^2 {\displaystyle\sum_{\{i,j,k\}}} G_R(p_i{+}p_j{+}p_k)
 	$.
	Here all 6 momenta involved in the two-to-four scattering are
	regarded as incoming, and
 	the sum runs over 10 distinct partitions of the six
	momenta into two groups of three momenta.
	Similarly, the sums
	in Eq.~(\protect\ref{eq:number_changing_scattering})
 	run over partitions of the five momenta into
	sets of 2 and 3 momenta, or 2, 2 and 1 momenta, respectively.
	}
 \begin{equation}
 {\cal T}_{\Delta N}
 =
 i \lambda \bar g \sum_{\{i,j\}} G_R(p_i{+}p_j)
 - i \bar g^3 \sum_{\{i,j\},\{l,m\}} G_R(p_i{+}p_j) G_R(p_l{+}p_m)
 \;,
 \label{eq:number_changing_scattering}
 \end{equation}
 where again $\bar g$ is the shifted cubic coupling constant.
 \item[{\it b})]
     The thermal mass is used in the dispersion relation for
     on-shell momenta, and in the retarded propagators appearing in the
     ``effective'' thermal scattering amplitudes
     (\protect{\ref{eq:almost_scattering_amp}}) and
     (\protect{\ref{eq:number_changing_scattering}}).
 \item[{\it c})]
    The physical mass (squared) appearing in the source term
    is replaced by
    \begin {equation}
	\tilde m^2 \equiv \mth^2 - T^2 \>
	{\partial \mth^2 \over \partial T^2} \,.
    \label {mtilde-def}
    \end {equation}
 \end{itemize}
    The ``subtracted'' mass $\tilde m^2$ is a measure of the
    departure from scale invariance.
    The subtraction 
    cancels the leading
    temperature dependence in $\mth^2$,
    so that $\tilde m^2$ differs negligibly from $\mphys^2$
    when $T \lsim \mphys/\sqrt\lambda$,
    and approaches
    $\mphys^2 - {\textstyle {1\over 2}} g^2/\lambda$
    for $\mphys/\sqrt\lambda \ll T \ll \mphys/\lambda$.
    At asymptotically large temperatures, $T \gg \mphys/\lambda$,
    the running of
    the quartic coupling in (\ref{eq:thermal_mass})
    dominates and $\tilde{m}^2 = \beta(\lambda) \, T^2/48$,
    up to $O(\sqrt{\lambda})$ corrections.
    %
    %
    %

 The resulting equation for the spin-0 response is
 \begin {eqnarray}
     {\textstyle{1\over 3}} {\bf p}^2
     - v_{\rm s}^2 \, ({\bf p}^2{+}\tilde m^2)
 &=&
     {E_p \over 2}
     \int_{123} d\Gamma_{12\leftrightarrow 3p}\,
     (1{+}n_1)\,(1{+}n_2)\,n_3 \, (1{+}n_p)^{-1}\,
 \nonumber
 \\
     && \qquad {}\times
     \bigg[
     A(p) + A(k_3) - A(k_2) - A(k_1)
     \bigg]
 \nonumber
 \\
 &+&
     {E_p \over 4}
     \int_{1234} d\Gamma_{12\leftrightarrow 34p}\,
     (1{+}n_1)\,(1{+}n_2)\,n_3 \, n_4\, (1{+}n_p)^{-1}\,
 \nonumber
 \\
     && \qquad {}\times
     \bigg[
     A(p) + A(k_4) + A(k_3) - A(k_2) - A(k_1)
     \bigg]
 \nonumber
 \\
 &+&
     {E_p \over 6}
     \int_{1234} d\Gamma_{123\leftrightarrow 4p}\,
     (1{+}n_1)\,(1{+}n_2)\,(1{+}n_3) \, n_4\, (1{+}n_p)^{-1}\,
 \nonumber
 \\
     && \qquad {}\times
     \bigg[
     A(p) + A(k_4) - A(k_3)  - A(k_2) - A(k_1)
     \bigg]
     \;,
 \label{eq:kinetic_bulk_eq_eff}
 \end {eqnarray}
 with
 \begin{equation}
     d\Gamma_{12\leftrightarrow 34p}
     \equiv
     {1\over 2 E_p}\,
     \Big| {\cal T}_{\Delta N}(p,k_4,k_3;k_2,k_1) \Big|^2 \,
     \prod_{i=1}^4 {d^3 {\bf k}_i \over (2\pi)^3 (2E_{k_i})}\,
     (2\pi)^4 \delta (P^{\rm tot}_{\rm in}{-}P^{\rm tot}_{\rm out}) \,,
 \label {diff-trans-rate-2}
 \end {equation}
 {\em etc}.
 In the pure quartic theory, the $2 \leftrightarrow 3$ particle
 terms are replaced by the corresponding $2 \leftrightarrow 4$
 particle contributions.
 Closing the effective vertex with an insertion of $\bar {\cal P}$
 yields the bulk viscosity,
 \begin {equation}
     \zeta = \beta
     \int {d^3 {\bf p}\over (2\pi)^3 E_p}\,
     \left(
     {\textstyle{1\over 3}}{\bf p}^2
     -
     v_{\rm s}^2 \, ({\bf p}^2 {+} \tilde m^2)
     \right)\,
     n(E_p)\, (1+n(E_p))\, A({\bf p}) \,,
 \label{eq:kinetic_bulk_eff}
 \end {equation}
 which differs from (\ref {eq:kinetic_bulk})
 by the replacement of $\mphys^2$ by $\tilde m^2$.%
 \footnote
 {%
 The solution of (\ref {eq:kinetic_bulk_eq_eff}) for $A(p)$
 is only unique
 up to the addition of a zero mode contribution proportional to $E_p$.
 This has no effect on the bulk viscosity (\ref {eq:kinetic_bulk_eff})
 because the speed of sound satisfies the identity
 \begin {equation}
     0 =
     \int {d^3 {\bf p} / (2\pi)^3}
     \left(
     {\textstyle{1\over 3}}{\bf p}^2
     -
     v_{\rm s}^2 \, ({\bf p}^2 {+} \tilde m^2)
     \right)\,
     n(E_p)\, (1+n(E_p)) \,.
 \label {speed-of-sound-identity}
 \end {equation}
 Nevertheless, the ambiguity in $A(p)$ may be eliminated by imposing the
 Landau-Lifshitz condition for the effective theory described below.
 This reduces to the constraint
 \begin {equation}
    0 =
     \int {d^3 {\bf p}\over (2\pi)^3 E_p}\,
     \left( {\bf p}^2 + \tilde m^2 \right)\,
     n(E_p)\, (1+n(E_p))\, A({\bf p}) \,.
 \label{eq:LL_constraint_eff}
 \end {equation}
 }

\section {Effective Kinetic Theory}

 Before discussing the solutions of these linearized
 equations for the hydrodynamic response,
 we wish to show how one may construct an
 effective kinetic theory for quasi-particle excitations
 which reproduces,
 at arbitrary temperature in a weakly coupled theory,
 the correct hydrodynamic response.
 %
 As usual, the quasi-particle distribution function $f(x,p)$ will depend on an
 on-shell four-momentum $p$, but now the quasi-particle
 energy $p^0 \equiv E_p$ will be a function of both the
 spatial momentum $\bf p$ and an effective mass $m(q)$,
 which in turn depends on a spacetime-dependent auxiliary field $q(x)$:
 \begin {equation}
    E_p(x) \equiv \left( {\bf p}^2 + m(q(x))^2 \right)^{1/2} \,.
 \label {eff-mass}
 \end {equation}
 The auxiliary field $q$
 characterizes the effect of the forward scattering
 of a quasi-particle off other excitations in the medium,
 and depends self-consistently on the distribution function,
 \begin {equation}
    q(x) \equiv \int {d^3p \over (2\pi)^3} \> { f(x,p) \over E_p(x) } \,.
 \end {equation}
 This is just a non-equilibrium generalization of the
 usual thermal contribution to the scalar field propagator at
 coincident points, $\langle \phi(x)^2 \rangle$.
 The quasi-particle Boltzmann equation can be written as
 \begin {equation}
    \left(
	 {\partial \over \partial t}
	 +
	 {\partial E_p \over \partial \bf p} \cdot
	 {\partial \over \partial \bf x}
	 -
	 {\partial E_p \over \partial \bf x} \cdot
	 {\partial \over \partial \bf p}
	 %
    \right)
     f(x,p)
     =
     \Delta \Gamma (x,p) \,.
 \end {equation}
 The dispersion relation (\ref {eff-mass}) implies that
 $\partial E_p / \partial {\bf p} = {\bf p} / E_p$
 and $\partial E_p / \partial {\bf x} = (m / E_p) \nabla m$.
 Hence, the spatial gradient of the effective mass acts like an
 external force which changes the momentum of propagating excitations.
 The collision term on the right hand side is the usual
 Boltzmann collision term with both $2 \leftrightarrow 2$
 and $2 \leftrightarrow 3$
 (or $2 \leftrightarrow 4$ for a pure quartic theory)
 particle processes included,
 \begin {eqnarray}
     \Delta \Gamma (x,p)
     &=&
     {\textstyle{1\over 2}}
     \int_{123} \> d\Gamma_{12\leftrightarrow 3p} \;\>
     \bigg(
     f_1\,f_2\,(1{+}f_3)\,(1{+}f_p)
     -
     (1{+}f_1)\,(1{+}f_2)\,f_3\,f_p
     \bigg) \!
 \nonumber
 \\ &+&
     {\textstyle{1\over 4}}
     \int_{1234} d\Gamma_{12\leftrightarrow 34p}
     \bigg(
     f_1\,f_2\,(1{+}f_3)\,(1{+}f_4)\,(1{+}f_p)
     -
     (1{+}f_1)\,(1{+}f_2)\,f_3\,f_4\,f_p
     \bigg)
 \nonumber
 \\ &+&
     {\textstyle{1\over 6}}
     \int_{1234} d\Gamma_{123\leftrightarrow 4p}
     \bigg(
     f_1\,f_2\,f_3\,(1{+}f_4)\,(1{+}f_p)
     -
     (1{+}f_1)\,(1{+}f_2)\,(1{+}f_3)\,f_4\,f_p
     \bigg) .
 \end {eqnarray}
 The transition rates (for a given spacetime location $x$)
 are given by the usual definitions (\ref {diff-trans-rate-1})
 and (\ref {diff-trans-rate-2}),
 with effective scattering amplitudes
 (\ref {eq:almost_scattering_amp}) and (\ref {eq:number_changing_scattering})
 computed using retarded free propagators containing the effective mass
 $m(q(x))$.

 This effective Boltzmann equation is to be combined with
 a modified definition of the kinetic theory stress-energy tensor,
 \begin {equation}
     T^{\mu\nu}(x)
     =
     \biggl(
     \int
     {d^3 {\bf p}\over (2\pi)^3 E_p} \>
     p^\mu p^\nu f(x,p)
     \biggr)
     -
     g^{\mu\nu} \, U(q(x))
     \,.
 \label {modified stress energy}
 \end {equation}
 A short exercise shows that the modified stress-energy
 tensor (\ref {modified stress energy}) is conserved
 provided the interaction energy $U(q)$ satisfies
 $
    \partial U / \partial q
    =
    -{\textstyle {1\over 2}} \, q \,
    (\partial m^2 / \partial q)
 $,
 or
 \begin {equation}
    U(q)
    =
    {\textstyle {1\over 2}}
    \int_0^q dq' \left( m^2(q') - m^2(q) \right) \,.
 \end {equation}
 This is also the necessary consistency condition for ensuring that
 the variation of the total energy density with respect to the
 quasi-particle density yields the correct quasi-particle energy,
 $
    E_p(x) = \delta \, T^{00}(x) \Big/ \delta f(x,p)
 $ \cite{Landau},
 and in equilibrium, that the pressure satisfy the
 correct thermodynamic identity
 $
    T (d {\cal P} / d T) = \varepsilon + {\cal P}
 $.

 The final ingredient needed to complete the definition of the
 effective kinetic theory is the dependence of the effective mass
 on the auxiliary field $q$.
 This is completely determined by the dependence of the
 equilibrium thermal mass $\mth$ on the one-loop ``bubble''
 $\langle \phi(x)^2 \rangle$.
 In the pure quartic scalar theory, the thermal mass
 has the simple form,
 \begin {equation}
    m^2(q) = m_0^2 + {\textstyle {1\over 2}} \, \lambda \, q
 \label {quartic_thermal_mass}
 \end {equation}
 (up to corrections suppressed by powers of $\lambda$),
 while if cubic interactions are present one must first self-consistently
 expand the field about its thermal expectation value
 $c \equiv \langle \phi \rangle$, leading to
 \begin {equation}
    m^2(q) = m_0^2 + g c + {\textstyle {1\over 2}} \, \lambda \, (c^2{+}q)
    \,,
 \label {cubic_thermal_mass}
 \end {equation}
 with
 $
    0 =
    m_0^2 \, c
    + {\textstyle {1\over 2}} \, g (c^2 {+} q)
    + {\textstyle {1\over 6}} \, \lambda \, c \, (c^2 {+} 3q)
 $.
As always,
the coupling constants appearing in
(\ref {quartic_thermal_mass}) and (\ref {cubic_thermal_mass})
should be evaluated at a scale appropriate
to the physics under consideration;
the running of the quartic coupling affects even leading order results
when $q \gsim \mphys^2/\lambda^2$.
In equilibrium, $q \sim T^2/12$ when $T \gg \mphys$.
Hence,
the appropriate generalization is to regard the coupling
as an implicit function of $q$ satisfying (when $q \gg \mphys^2$)
\begin {equation}
    q \, {\partial \lambda \over \partial q}
    \equiv {\textstyle {1 \over 2}} \, \beta (\lambda)
    = {\textstyle {1 \over 2}} \, b_0 \, \lambda^2 + O(\lambda^3)
    \,.
\end {equation}
The resulting effective mass in, for example, the massless pure
quartic theory is
\begin {equation}
    m^2(q)
    = {q \over b_0 \ln (\Lambda^2/q)} \,,
\end {equation}
where $\Lambda \equiv \mu \, e^{1/b_0 \lambda(\mu^2)}$
is the renormalization group invariant scale of massless $\phi^4$ theory.

 This effective kinetic theory provides a consistent description
 of the non-equilibrium dynamics of a weakly coupled scalar field theory,
 including the propagation of slowly moving excitations,
 even when the effective mass of the excitations differs substantially
 from the zero-temperature mass, or varies significantly in space or time.
 Expanding about a local equilibrium distribution,
 as in (\ref {eq:f_plus_df}),
 and evaluating the effective stress energy tensor
 (\ref {modified stress energy})
 (carefully keeping track of the implicit dependence on the
 distribution function hiding in every factor of energy),
 leads to the fairly simple result
 \begin {eqnarray}
    T^{\mu\nu}(x)
    &=&
    T^{\mu\nu}_{\rm eq}(x) -
    \int {d^3p \over (2\pi)^3 E_p} \>
    n(E_p) (1{+}n(E_p)) \, \chi(x,p)
    \biggl(
	p^\mu p^\nu - u^\mu u^\nu \> T^2 \, {\partial m^2 \over \partial T^2}
    \biggr),
 \label {eff-linearized-stress}
 \end {eqnarray}
 where
 $
     T^{\mu\nu}_{\rm eq} \equiv
     u^\mu u^\nu \, (\varepsilon {+} {\cal P})
     + g^{\mu\nu} \, {\cal P}
 $
 is the local-equilibrium contribution.
 Expressing $\chi(x,p)$ in terms of the shear and bulk amplitudes
 ({\em c.f.} Eq.~(\ref{phi-from-A&B})), and
 linearizing the effective Boltzmann equation in the hydrodynamic limit,
 yields exactly the same equations obtained in the previous section
 for the amplitudes $A(x,p)$ and $B(x,p)$.
 When inserted into the stress tensor (\ref {eff-linearized-stress})
 one precisely obtains the previous results
 (\ref{eq:kinetic_shear}) and (\ref{eq:kinetic_bulk_eff})
 for the shear and bulk viscosities.


\section {Results for Viscosities}\label{results}

 Computing the bulk viscosity requires solving the
 integral equation (\ref{eq:kinetic_bulk_eq_eff}).
 Unlike the case of the shear viscosity, solving this equation
 is trivial because the kernel has a single small eigenvalue
 which is only displaced from zero due the inclusion of number changing
 processes.  Hence, the solution is dominated by the projection onto the
 near-zero mode, leading to
 \begin{equation}
 A(p) = {F \over \Gamma_{\Delta N}} \> (1 - \alpha \, E_p)
 \;,
 \label {eq:A_soln}
 \end{equation}
 where
 \begin{equation}
     F
     \equiv
     \int
     {d^3 {\bf p} \over (2\pi)^3 E_p}\,
     [1{+}n(E_p)]\,n(E_p)\,
     I_{\bar{\cal P}}(p)
     \;,
     \label{eq:bI_p}
 \end{equation}
 with
 $
     I_{\bar{\cal P}}(p)
     \equiv
     {\textstyle{1\over 3}}{\bf p}^2
     {-}
     v_{\rm s}^2 ({\bf p}^2 {+} \tilde m^2)
 $
 the same source term as
 in Eq.~(\ref{eq:kinetic_bulk_eq_eff}),
 and $\Gamma_{\Delta N}$ the total $3\to 2$ particle
 (or $4\to 2$ for pure $\lambda\phi^4$)
 thermal reaction rate per unit volume,
 \begin{eqnarray}
 \Gamma_{\Delta N}
 &=& \displaystyle
 {1\over 12}
 \int
 \prod_{i=1}^5
 {d^3 {\bf k}_i \over (2\pi)^3 2 E_{k_i}}\,
 \Big| {\cal T}_{\Delta N}(\{k_i\}) \Big|^2\,
 (2\pi)^4\delta(k_1{+}k_2{+}k_3{-}k_4{-}k_5)\,
 \nonumber\\
 & & \displaystyle \qquad {}\times
 \bigg(
 [1{+}n(E_1)]\, [1{+}n(E_2)]\,
 n(E_3)\, n(E_4)\, n(E_5)\,
 \bigg)
 \;.
 \label{eq:Gamma_dN}
 \end{eqnarray}
 The constant $\alpha$ in (\ref {eq:A_soln}) is undetermined by
 (\ref {eq:kinetic_bulk_eq_eff}), but may be adjusted to satisfy
 (\ref{eq:LL_constraint_eff}).
 %
 %
 %
 The bulk viscosity
 obtained by inserting (\ref {eq:A_soln}) into (\ref {eq:kinetic_bulk_eff})
 is simply
 \begin{equation}
 \zeta
 =
 \beta \,
 {
         F^2
	\over
 \Gamma_{\Delta N}
 }
 \;.
 \label{eq:leading_bulk}
 \end{equation}

 The final evaluation of the shear viscosity requires a numerical
 solution of the integral equation (\ref{eq:kinetic_shear_eq})
 and the final integral (\ref{eq:kinetic_shear}), while the bulk
 viscosity requires performing the rather involved phase space integral
 (\ref{eq:Gamma_dN}) for the particle number changing reaction rate.
 Details of this evaluation may be found in \cite{jeon2}.

 Despite the need to resum self-energy insertions in order to cut off
 singularities in the original diagrams, the introduction of thermal
 corrections in the dispersion relation and scattering amplitude is
 actually irrelevant for the leading behavior of the shear viscosity
 because the integrals (\ref{eq:kinetic_shear_eq}) and
 (\ref{eq:kinetic_shear}) are dominated by momenta of order $T$.  This,
 however, is not the case for the bulk viscosity.  At high temperature,
 the number changing reaction rate scales as
 $O(g^2\lambda^2 T^5/\mth^3)$
 for the $g\phi^3{+}\lambda\phi^4$ theory, and
 $O(\lambda^4 T^6/\mth^2)$ for the pure $\lambda\phi^4$ theory
 due to its infrared sensitivity to the region where all momenta are
 $O(\mth)$.  The factor $F$ appearing in the
 numerator is a measure of the violation of scale invariance of the
 theory, and behaves as $O(\tilde m^2 T^2\ln(T/\mth) )$ when
 $\mphys \ll T$.%
 \footnote
	{%
	One finds
	$
	    v_{\rm s}^2 =
	    {\textstyle {1\over 3}} -
	    {\textstyle {5\over 12}} \, \tilde m^2 / \pi^2 T^2
	$
	and
	$
	    F = -(\tilde m^2 T^2/6\pi^2)
	    \left[\,
		\ln(2T/\mth) - {\textstyle {15\over 2}} \zeta(3)/\pi^2
	    \,\right]
	$,
	when evaluating (\ref {speed-of-sound-identity}) and
	(\ref{eq:bI_p}) for $T \gg \mphys$,
	up to corrections suppressed by $(\sqrt \lambda)$ or $\mphys/T$.
	}
 Hence, the shear and bulk viscosities have very different behaviors
 throughout the high temperature region.\footnote
    {%
    They are also very different at low temperature.
    When $T \ll \mphys$,
    the shear viscosity behaves like
    $\eta \sim \mphys^3(T/\mphys)^{1/2}/\lambda^2$,
    but the bulk viscosity
    diverges exponentially as
    $\zeta \sim e^{2\mphys/T} \, \mphys^6/\lambda^4 T^3$
    for pure $\lambda\phi^4$ theory, or
    $e^{\mphys/T} \, (\mphys/T)^{1/2} \,\mphys^6/\lambda^2 g^2 T$
    for $g\phi^3{+}\lambda\phi^4$ theory.
    This is the bulk viscosity characterizing asymptotically long wavelength
    hydrodynamic fluctuations, appropriate for distances large compared
    to the mean free path for particle number changing interactions
    (which displays the same exponential divergence).
    Ordinary non-relativistic hydrodynamics (with a conserved particle
    number) is valid at distances small compared to this number changing
    mean free path but large compared to the elastic mean free path.
    It is, of course, this region and not the strict asymptotic domain
    which has practical utility.
    }
 In pure $\lambda\phi^4$ theory,
 \begin{equation}
 \eta = a \, {T^3\over \lambda^2} \times
 \left[\, 1 + O(\sqrt{\lambda}) + O(\mphys/T) \,\right] ,
 \label{eq:4_shear}
 \end{equation}
 while
 \begin{equation}
 \zeta =
    b \, {\tilde m^4 \mth^2 \over \lambda^4 T^3} \,
    \ln^2\Bigl({\kappa \,\mth \over T } \Bigr)
    \times \left[\,
    1 + O(\sqrt{\lambda}) + O(\mphys/T) + O(\lambda T/\mphys)
    \,\right] ,
 \label{eq:4_bulk_mid}
 \end{equation}
 when $\mphys \ll T \ll \mphys/\lambda$, and
 \begin{equation}
 \zeta = c \, \lambda \ln^2({\gamma\,\lambda}) \, T^3
 \times \left[\,
 1 + O(\sqrt{\lambda}) + O(\mphys/\lambda T)
 \,\right] ,\kern 40pt
 \label{eq:4_bulk_high}
 \end{equation}
 when $T \gg \mphys/\lambda$.
 The forms (\ref{eq:4_shear}) and (\ref{eq:4_bulk_high}) remain valid
 if cubic interactions are present, but the bulk viscosity in the
 intermediate regime $\mphys \ll T \ll \mphys/\lambda$ acquires
 dependence on the relative strength of cubic and quartic
 couplings,
 \begin{equation}
 \zeta =
    d\Bigl({g^2 \over \lambda \mth^2}\Bigr)
    {\tilde m^4 \, \mth^3 \over g^2 \lambda^2 T^2} \,
    \ln^2\Bigl({\kappa \, \mth \over T } \Bigr) \!
    \times \left[\,
    1 + O(\sqrt{\lambda}) + O(\mphys/T) + O(\lambda T/\mphys)
    \,\right] ,
 \label{eq:3_bulk_mid}
 \end{equation}
 with $d(x)$ a non-trivial dimensionless function.

\begin{figure}
\setlength {\unitlength}{1cm}
\vbox
    {%
    \begin {center}
	\begin{picture}(0,0)
	    \put(-1.5,4.0){$\log_{10}\left({\eta\lambda^2\over \mth^3} \right)$}
	    \put(5.75,-0.5){${T / \mth}$}
	\end{picture}
	\leavevmode
	\def\epsfsize  #1#2{0.45#1}
	\epsfbox {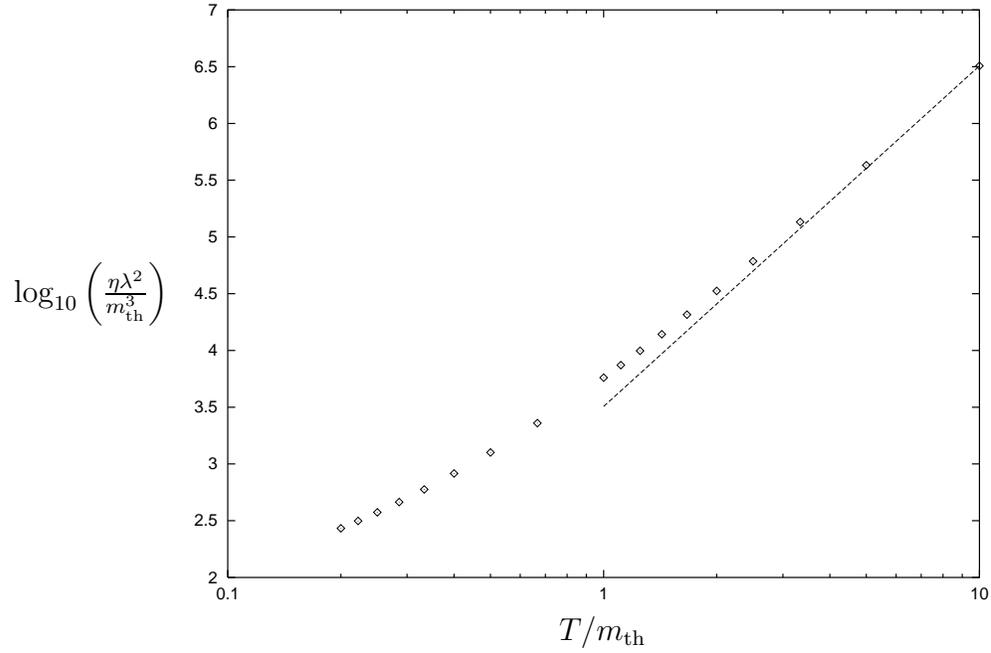}
    \end {center}
    \vspace*{1.0cm}
    \caption
    {%
    Numerical results for the shear viscosity.
    The straight line shows the $O(T^3/\lambda^2)$
    asymptotic behavior of $\eta$.
    }
    \label{fig:shear_fit}
    }
\end{figure}

\begin{figure}
\setlength {\unitlength}{1cm}
\vbox
    {%
    \begin {center}
    \begin{picture}(0,0)
       \put(-1.8,4.0)%
	   {$\log_{10}\left({\zeta\lambda^4\mth\over\tilde m^4}\right)$}
	\put(5.75,-0.5){${T / \mth}$}
    \end{picture}
    \leavevmode
    \def\epsfsize  #1#2{0.45#1}
    \epsfbox {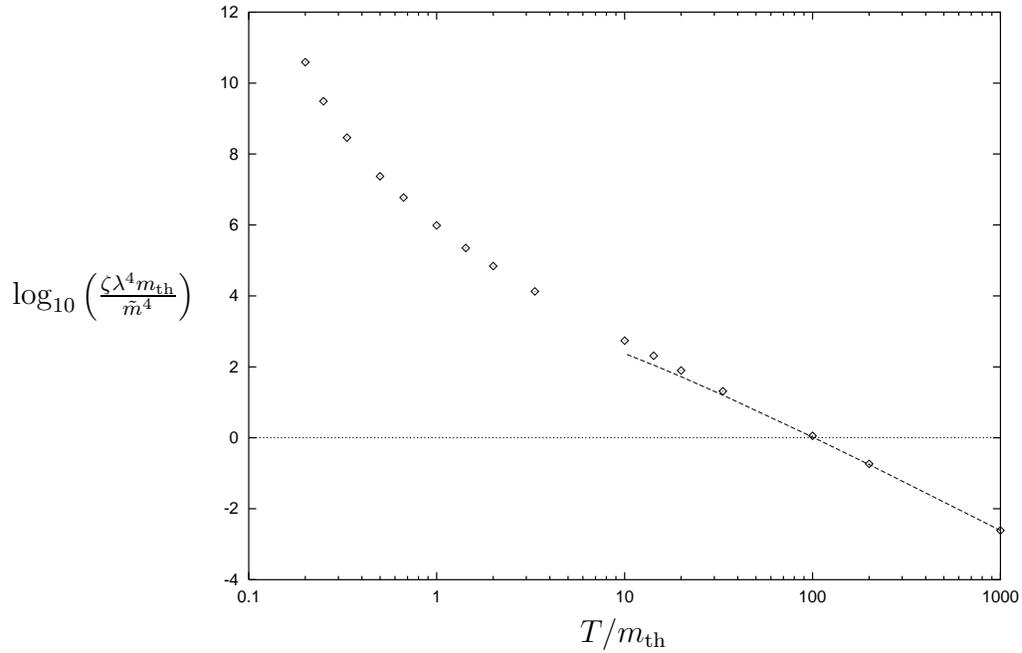}
    \end {center}
    \vspace*{1.0cm}
    \caption
    {%
    Numerical results for the bulk viscosity.
    The solid line shows the $(\mth/T)^3 \, \ln^2(\kappa \, \mth/T)$ behavior
    of Eq.~(\protect\ref {eq:4_bulk_mid}).
    }
    \label{fig:bulk_fit}
    }
\end{figure}

 A numerical evaluation of equation (\ref{eq:kinetic_shear_eq}),
 (\ref{eq:kinetic_shear}), (\ref{eq:Gamma_dN}), and
 (\ref{eq:leading_bulk}) for the pure quartic theory
 yields the values \cite{jeon2}:%
 \footnote{%
	The result (\protect{\ref{eq:numbers}}c)
	was not included in \cite{jeon2}.
	In addition, the evaluation of $\Gamma_{\Delta N}$ in ref.~\cite{jeon2}
	contained a numerical error which affected the plot
	of $\zeta$ shown in that paper.
	Recomputed values have been used in our Fig.~\protect\ref {fig:bulk_fit}
	and Eq.~(\protect\ref{eq:numbers}b).
	}
 \begin{mathletters}
 \begin{eqnarray}
 a &=& \displaystyle 3.04 \times 10^3 \,,
 \\
 b &=& \displaystyle 5.5\times 10^4 \,,
 \\
 c &=& 
	 b / \bigl(6 \, (32 \pi)^4 \bigr)
     = \displaystyle 8.9\times 10^{-5}  \,,
 \\
 \kappa &=& \displaystyle e^{15\zeta(3)/2\pi^2}/2 = 1.2465 \,,
 \\
 \gamma &=& \displaystyle e^{15\zeta(3)/\pi^2}/96 = 0.064736 \>.
 \end{eqnarray}
 \label{eq:numbers}%
 \end{mathletters}%
 \noindent
 Results in the relativistic cross-over region $T \sim \mth$
 are plotted in Figs.~\ref {fig:shear_fit} and \ref {fig:bulk_fit}.
 If one ignores the need to sum all ladder diagrams and only
 includes the one-loop diagram of Fig.~\ref{fig:one_loop}
 (after resumming self-energy corrections) then one underestimates the shear
 viscosity by roughly a factor of four.  The analogous error for the
 bulk viscosity leads to an $O(\mphys^4/\lambda^2 T)$ result which
 scales completely incorrectly with $\lambda$.

\section {Conclusions}

    The analysis of this simple scalar field theory illustrates
 a number of points which are applicable to any relativistic field theory:
\begin{itemize}
\item[{\it a})]  The diagrammatic evaluation of transport coefficients
	is a remarkably inefficient approach.
	An infinite set of rather complicated diagrams must be summed,
	merely to obtain the leading weak coupling behavior.

\item[{\it b})]	The bulk viscosity depends on particle number changing
	processes and is sensitive to soft momenta,
	whereas the shear viscosity is determined by two body
	elastic scattering cross sections at typical momenta.
	The ratio of the bulk to the shear viscosities varies
	from very small ($O(\lambda^3)$) to exponentially large
	depending on the temperature.
	Hence, crude estimates such as
	$\zeta \sim \eta \, (v_{\rm s}^2 {-} {\textstyle {1 \over 3}})^2$
	which have appeared in the literature \cite{Hosoya,Horsley}
	cannot generally be trusted.


\item[{\it c})]	At high temperature, the existence of an effective kinetic
	theory adequate for computing transport coefficients
	depends crucially on the theory being weakly coupled,
	so that mean free paths are large compared to the
	wavelengths of relevant excitations.
	In, for example, high temperature QCD,
	it is unclear if the bulk viscosity can be correctly
	computed with any kinetic theory since the effective
	coupling of excitations with soft $O(g^2 T)$ momenta
	is not small.
\end{itemize}

    It is tempting to view
    the derivation of kinetic theory from the underlying field theory,
    and the derivation of the hydrodynamic constitutive equation (1.2)
    from the effective kinetic theory,
    as two different stages of a ``real time renormalization group''.
    At each stage, one is eliminating irrelevant degrees of freedom
    from the description of dynamics at successively lower frequency
    or momentum scales.
    We have little doubt that this notion of a real time renormalization
    group is essentially correct.
    However, we are unaware of any useful framework for defining
    a real time renormalization group which will systematically
    transform the basic dynamical formulation from a quantum field theory
    to kinetic theory, or ultimately to classical hydrodynamics.
    In contrast to the situation for equilibrium Euclidean space
    observables \cite{Braaten},
    how to repackage the cumbersome diagrammatic analysis of~\cite{jeon2}
    in simple renormalization group terms is poorly understood.
    The diagrammatic treatment does not cleanly separate different
    frequency scales, as shown, for example, by the necessity of
    resumming both the real and imaginary parts of the on-shell
    single particle self energy in order to regulate
    individual cut diagrams, even though only the real part
    of the self energy appears explicitly in the resulting kinetic theory.
    The imaginary self energy, or single particle lifetime,
    should be viewed as an output of the effective kinetic theory, not an
    input parameter.
    A true real time renormalization group approach should allow
    one to derive completely the effective kinetic theory
    before treating any of the physics for which the kinetic theory
    description is adequate.
    Furthermore, a useful renormalization group framework
    should allow one to calculate corrections systematically,
    at least in weakly coupled theories.
    Although an effective kinetic theory did emerge
    in the analysis of the leading weak coupling behavior,
    it is unclear whether subleading corrections can be incorporated within
    a kinetic theory framework, since quantum coherence effects
    are only suppressed by a power of $\lambda$.
    We hope that future investigations will shed light on some
    of these issues.

\acknowledgements{%
    Helpful conversations with Peter Arnold and Lowell Brown are
    gratefully acknowledged.
}

\end{document}